\documentstyle[prl,preprint,aps]{revtex}
\begin{document}

\draft

\title{Theory of the Eigler-switch}
\author{Mads Brandbyge and Per Hedeg\aa{}rd}
\address{
\O{}rsted Laboratory, Niels Bohr Institute,\\
Universitetsparken 5, DK-2100 Copenhagen, Denmark
}
\date{\today}
\maketitle
\begin{abstract}
We suggest a simple model to describe
the reversible field-induced transfer of a single Xe-atom in a scanning
tunneling microscope, --- the Eigler-switch. The
inelasticly tunneling electrons give rise to fluctuating forces on
and damping of the
Xe-atom resulting in an effective current dependent temperature.
The rate of transfer is controlled by the well-known Arrhenius law with this
effective temperature. The directionality of atom transfer is discussed,
and the importance of use of non-equlibrium-formalism for the electronic
environment is emphasized. The theory constitutes a formal derivation and
generalization of the so-called Desorption Induced by Multiple Electron
Transitions (DIMET) point of view.
\end{abstract}

   Eigler and Schweizer\cite{eig90} have been able to manipulate
Xe atoms and place them with atomic precision on a Ni surface using an
ultra high vacuum scanning tunneling microscope (STM) operated at
4 Kelvin.
Experiments by Eigler, Lutz and Rudge show\cite{eig91},
that a single Xe atom physisorbed to a particular kink site on a
single crystal Ni-(110) surface,
can be transferred reversibly between this surface-site and the W tip in
the STM.
These experiments are done at
4 Kelvin. Using voltage pulses of $\pm 0.8V$ for $64 msec$, they
are able to toggle the Xe atom from surface to tip and
back to the same position on the surface at will, and find that the
direction of transfer is the {\em same} as that of the tunneling electrons.
For a particular tip-surface configuration with the Xe adatom on the
surface,
the transfer rate, $\tau^{-1}$, goes as $I^{4.9\pm0.2}$, $I$
being the tunneling current. The voltage range in these measurements is from
$18mV$ to $180mV$ with $V/I=906k\Omega\pm2\%$.
These phenomena has been investigated theoretically
by Walkup, Newns and Avouris
\cite{newns} and Gao, Persson and Lundquist\cite{gao}. These authors suggest
that the main mechanism behind the transfer is that the current
excite the Xe atom vibrationally
in the double-well potential, sustained by the van der
Waal attraction to surface and tip. On the other hand the Xe atom dissipates
energy to the surface phonons, so they calculate the rate of transfer using
rate-equations (Pauli master equations) including the competition between
dissipation to surface-phonons and ``heating'' by inelasticly tunneling
electrons.
In Ref. \cite{newns} the
possible mehanisms behind the directionality of transfer is
discussed, and they conclude that
the adsorbtion-induced dipole is the dominant effect.

In this paper we concentrate on showing that the key-features of
the experiments can be explained from a less phenomenological point of view,
based on a simple model.

    The main idea in our calculation is the following.
We view the Xe atom as a quantum Brownian particle interacting with the
environment of electrons in the tip and surface via the
adsorbate-resonance, through which the electrons tunnel. Besides this
environment the Xe atom also interact with an environment of surface-phonons.
The influence of these environments on the atom is described in a
path-integral framework using the
influence-functional introduced by Feynman and
Vernon\cite{feynmanvernon,feynhibbs}, giving an
effective action describing the motion of the atom.
For two independent environments,
the influence-functional will simply be the product of
the individual influence-funtionals\cite{feynhibbs}. The influence-functional
for a harmonic oscillator with a linear coupling to a
continuous distribution of harmonic oscillators in thermal
equilibrium has been calculated by Caldeira and Legget\cite{callegg}.
This influence-functional contains the fluctuating force and dissipation
caused by the environment which in the classical limit are the
ingredients in
the Langevin equation describing the classical motion\cite{schmid}.
Our contribution is the calculation of the influence-functional
corresponding to the non-equilibrium electronic environment.
In general,
when a Brownian particle is in equilibrium with a heat-bath, the
fluctuating force acting
on the particle and the corresponding friction force are ``connected'' by the
fluctuation-dissipation theorem (FDT).
This is the situation when no chemical-potential difference between
tip and surface drives a current through the adsorbate-resonanse, and the
Xe atom is in equilibrium, the surface-phonons and the electrons
acting as heat baths. When an external field in
some way is transferring energy to the Brownian particle the
relation between the fluctuating force and the dissipation, found
in the equilibrium case, no longer holds. There will in general
occur a extra current-dependent fluctuating force not
``compensated'' by friction.

We show this
general feature in our model, where the interaction
with the inelasticly tunneling electrons give rise to
the uncompensated fluctuating
force. At sufficiently low temperature compared with the oscillation frequency
in the adsorbtion well, as in the actual experiment,
this force will be the dominant. The
temperature independent dissipation is due to exitation of
surface-phonons and electron-hole-pairs in the surface and tip by the
vibrating Xe atom.
The time scale of desorbtion is much longer than
than the relaxation time of the dissipative system, thus we have a
quasi-stationary situation a relaxation time after the current is turned on.
In this quasi-stationary situation we can consider the Xe-atom
as a damped harmonic oscillator with a wavepacket width
given by the usual FDT-result plus a contribution due to the non-equilibrium
fluctuating force. From this width we define our effective temperature.
The desorbtion rate is then given by the well-known results for
escape of a Brownian particle from a potential well in equilibrium with
a heat-bath where the new effective temperature depends on the voltage.
This leads to the experimentally found result if we assume
that the adsorbtion well contains
about five bound states, as was done in
Refs. \cite{newns} and \cite{gao} .
The dependence of the direction of transfer on the polarity of field
follows in our model from the mean occupation of the Xe atom.
We stress that it is
crucial to use a non-equilibrium formalism to calculate the
occupation.

Let us now introduce the model and sketch the derivation of our
results.
The reason why a rare-gas atom like Xe actually can be imaged by STM is
explained by Eigler {\em et al.} \cite{lang}.
Thus we will only consider this adsorbate state.
Denoting the operator and energy associated with the Xe 6$s$-orbital $d$
and $\epsilon_a$, and the
operators asociated with surface, respectively tip, $c_k$, $c_l$, we
write the Hamiltonian for the electronic part,
\begin{equation}\label{hamiltonian}
H_{el} = H_{surf} + H_{tip} + \epsilon_a d^{\dagger} d +
\sum_{k,l} d^{\dagger}(T_{k}c_{k}+T_{l}c_{l}) + h.c.
\end{equation}
We define the weighted density of states for
surface and tip,
\begin{eqnarray}
\rho_s(E)&=& \sum_{k}\;\vert T_{k}\vert^2
\delta(E-\epsilon_{k})                          \\
\rho_t(E)&=& \sum_{l}\;\vert T_{l}\vert^2
\delta(E-\epsilon_{l}).
\end{eqnarray}
For simplicity we use the wideband limit and take these to be constant.
The current, corresponding to the model above,
can be expressed by a non-equilibrium
Green's function which is evaluated
using the Keldysh technique
\cite{smithrammer}, yielding
\begin{eqnarray}
\lefteqn{I(V)=}\\
&&\frac{4e}{\hbar}\int_{-\infty}^{\infty}\frac{d\omega}{2\pi}
 \frac{(\pi \rho_t)(\pi \rho_s)(n_F(\omega-eV)-n_F(\omega))}
{(\pi(\rho_t+\rho_s))^2+(\omega-\epsilon_a)^2}.
\nonumber
\end{eqnarray}
We have neglected the small influence of the vibrations of
the Xe atom on the current.

The parameters of the electronic Hamiltonian (\ref{hamiltonian}) depends
on the position of the adatom. This of course leads to the picture of
effective potential energy-surfaces, which is the basis of most surface
dynamics calculations. We shall see that this dependence also is responsible
for the fluctuations in the force experienced by the atom. Denoting by
$\hat{x}$ the operator associated with the position of the Xe-atom, relative
to the equilibrium distance from the surface, and expanding the the
onsite energy to first order in $\hat{x}$ we obtain the coupling between
electrons and Xe
\begin{equation}
V_{el-vib}=g (d^{\dagger}d-\langle d^{\dagger} d\rangle_{eq})\;\hat x,
\end{equation}
where g is the {\em positive} coupling-constant. (For simplicity we
neglect the $\hat{x}$-dependence of $T_k$ and $T_l$).
Note that we have subtracted
the equilibrium ($eV=0$) mean population of the site, because this
contribution is assumed already to be incorporated in the adsorbtion
potential-surface. The physical picture of the interaction is that
the temporary increase in charge on the site, when a electron tunnel through
it, will cause the atom to be attracted more
strongly towards the strongest image, which will be the one in the surface
when the atom is located at the surface. Thus $g$ will be positive.
The interaction results in a fluctuating force acting
on the Xe atom, due to the electrons, jumping on and off the
site. This physical effect is what has been dubbed DIMET (desorbtion
induced by multiple electron transitions)\cite{dimet}.
On the other hand,
the vibration of the atom will, because of this interaction,
excite electron-hole pairs in the
metal-surface and dissipate energy to the electronic part of the system.
A similar model has been used by Persson and Baratoff\cite{baratoff} to
describe the influence of inelastic tunneling on the current in the STM.

We use the Feynman-Vernon theory of influence-functionals
to do a systematic calculation of these features.
Considering a quantum system interacting with a general environment, one
is often only interested in the influence of the environment on the system.
This can be accomplished by the reduced density operator, tracing out
all superfluous information of the environment. This is elegantly done in
a path-integral framework\cite{feynhibbs}. The result of this procedure
is an effective action of the system.
This approach has been used on the problem of
a quantum oscillator in equilibrium with a
heat-bath consisting of a continuum of
quantum oscillators, and is treated thoroughly in Refs.
\cite{feynmanvernon,feynhibbs,callegg,schmid}.
In our problem we will be concerned only with the behavior of the
Xe-oscillator and thus trace out the
equilibrium phonon environment as well as the non-equilibrium
electronic environment.

The propagator for the reduced density-operator for the oscillator can
be written as a double path integral,
\begin{eqnarray}
\lefteqn{J(x_f,y_f,t;x_i,y_i,0)=}\\
&&\int{\cal D}x\int{\cal D}y\;e^{i (S(x)-
S(y))/\hbar} {\cal F}_{el}(x,y){\cal F}_{ph}(x,y),\nonumber
\end{eqnarray}
where $S$ is the action for the non-interacting oscillator, and
${\cal F}_{el}$ and ${\cal F}_{ph}$ are
the influence-functionals containing all relevant information of the
electronic and phonon environments, respectively. These will, in general,
couple the
paths moving ``forward'' ($x$) and ``backward'' ($y$) in time. The
phonon influence
functional, corresponding to the oscillator coupled to a continuum of
oscillators, has been evaluated in Ref. \cite{callegg},
\begin{eqnarray} \label{eq:phon}
\lefteqn{\Delta S_{\text{\it eff}}^{ph}(x(\tau),y(\tau))=-i\hbar\:
\ln[{\cal F}_{ph}(x,y)]}\nonumber \\
&=&-\frac{\eta_{ph}}{2}\int_{0}^{t} (\dot x + \dot y)(x-y)d\tau \\
&&+i \eta_{ph}\int_{0}^{\infty}\omega
\coth\left(\frac{\hbar\omega}{2kT}\right)  \nonumber \\
&&\int_{0}^{t}\int_{0}^{t}(x(\tau)-y(\tau))\cos(\omega(\tau-s))(x(s)-y(s))
d\tau\;ds\frac{d\omega}{2\pi}. \nonumber
\end{eqnarray}
The real part of $S_{\text{\it eff}}$ describes the friction\cite{langevin} and
the imaginary
is the correlation function of the time-non-local
fluctuating force. These two terms are related through the equilibrium FDT.

The new feature here is the contribution to $S_{\text{\it eff}}$
from the non-equilibrium electronic environment.
This is calculated using a technique similar to the one
used in Ref. \cite{hedeg}, extended to cover this non-equilibrium case.
The first order term is,
\begin{eqnarray}
\lefteqn{\Delta S_{\text{\it eff}}^{el,(1)}(x,y)=}\\
&&-\;g\;(N(eV)-N(0))\:\int_{0}^{t}dt'\;(x(t')-y(t')),\nonumber
\end{eqnarray}
where $N(eV)$ is the mean population of the Xe site after the
voltage is applied, and is given by,
\begin{eqnarray}
\lefteqn{N(eV)=} \\
&&\int_{-\infty}^{\infty}d\omega
\:\frac{\rho_t\;n_F(\omega-eV)\:+\:\rho_s\;n_F(\omega)}
{(\omega-\epsilon_a)^2+
\pi^2(\rho_t+\rho_s)^2},\nonumber
\end{eqnarray}
where the $\epsilon_F^{surf}=0$ and $\epsilon_F^{tip}=eV$.
This corresponds to a simple
change of the potential energy surface due to the change of population of the
site. The mean-population of the resonance
increases for $eV$
positive, and thus increase the energy-barrier between the surface and
tip adsorbtion-sites, whenever the electrons tunnel from tip to surface.
Reversing
the polarity, so the electrons tunnel from surface to tip,
decreases the barrier. When the Xe-atom is located on the tip, the roles
played by tip and surface are exchanged. This can explain the observed
dependence of transfer on the polarity.

To second order in the coupling constant $g$
we find terms similar to (\ref{eq:phon}) with a friction-coefficient,
$\eta_{el}(eV)$, which can be expressed in terms of $g$, $\rho_s$, $\rho_t$ and
$eV$\cite{unpub}. Besides these terms we find the additional imaginary term
in $\Delta S_{\text{\it eff}}^{el,(2)}$,
\begin{eqnarray}
\lefteqn{\alpha(eV)\int_{0}^{t}\int_{0}^{t}
\int_{-\infty}^{\infty}
\;[\coth(\frac{(\hbar\omega-eV)}{2kT})-\coth(\frac{\hbar\omega}{2kT})]}
\nonumber \\
& &(x(\tau)-y(\tau))\cos(\omega(\tau-s))(x(s)-y(s))
\;d\tau\;ds\;\frac{d\omega}{2\pi},
\end{eqnarray}
with
\begin{equation}
\alpha(eV)=-\frac{g^2}{2}
\int_{0}^{eV}
\frac{\rho_s \rho_t}
{[(\omega-\epsilon_a)^2+
\pi^2(\rho_t+\rho_s)^2]^2}\;d\omega.
\end{equation}

The effective action obtained so far is quadratic in the Xe-atom coordinate,
and the remaining path-integral is Gaussian and can be worked out exactly.

The width of a wave-packet describing
the spatial motion of a damped harmonic oscillator with frequency
$\omega_0$ and mass $M$ in equilibrium
with a heatbath is determined by the imaginary part of the response function
for damped oscillator $\chi''(\nu)$ through the FDT,
\begin{eqnarray}
\sigma^2&=&\frac{\hbar}{\pi}\int_{0}^{\infty}\coth\left(
\frac{\hbar \nu}{2kT}\right)\;
\chi''(\nu)\\
\chi''(\nu)&=&\frac{1}{M}\frac{\left (\frac{\eta\nu}{M}\right )}
{(\nu^2-\omega_0^2)^2 + (\frac{\eta \nu}{M})^2}.
\end{eqnarray}
{}From the discussion by Caldeira and Legget \cite{callegg}, it follows
immediatly that this also will be the case in equilibrium ($eV=0$)
for our model, with $\eta = \eta_{ph} + \eta_{el}$.
For sufficiently low damping, where the energy eigenstates of the
oscillator is welldefined, $\chi''(\nu)$ will be a narrow function
centred around $\omega_0$, and for low temperature we get,
\begin{equation}
\sigma^2\propto 1+2e^{-\hbar\omega_0/kT}.
\label{eq:noalfa}
\end{equation}
Carrying out a similar analysis in our non-equilibrium case, we find in
the same limit, that
\begin{equation}
\sigma^2\propto 1+2e^{-\hbar\omega_0/kT}+
\frac{\vert \alpha(eV) \vert}{\eta_{tot}(eV)\omega_0} .
\label{eq:alfa}
\end{equation}
Comparing the non-equilibrium expression (\ref{eq:alfa}) with the
equilibrium expression (\ref{eq:noalfa})
enable us to define an effective temperature
by equating the small terms:
\begin{equation}
\exp\left(-\frac{\hbar\omega_0}{kT_{\text{\it eff}}(eV)}\right)
=e^{-\hbar\omega_0/kT}+
\frac{\vert \alpha(eV)\vert}{2\eta_{tot}(eV)\omega_0}.
\end{equation}
In figure 1 we have plotted the effective temperature as a function
of the voltage $eV$.

In the case where we can neglect the exponentially small
$e^{-\hbar\omega_0/kT}$, the effective Boltzmann factor is
proportional to $\alpha$ and thus to $eV$.
The observed rate can now be understood from the Arrhenius-factor
$e^{-U_0/kT}$ dominating the rate-expressions of escape problems in
general\cite{melnikov}, where $U_0$ is the barrier-height. If we,
as suggested in \cite{newns,gao}, assume that
$U_0\approx 5\hbar\omega_0$, we conclude that $\tau^{-1}\propto (eV)^5$.
In figure 2 the escape rate is plotted as a function of voltage for
different temperatures.
We predict a drastic change in this behavior as the temperature is raised
which can be verified experimentally.

We have not, in this calculation, taken into account
the dependence of the tunnelmatrixelements
on the adsorbate position ($T_{k}(x)$, $T_{l}(x)$). This will contribute to
the friction as well as the fluctuating forces on the Xe-atom.
Taking this dependence into account also, the
general scheme described above can be used
to calculate the friction coefficient from a microscopic starting point,
using given expressions for
$\epsilon_a(x)$ and $T_{k}(x)$, $T_{l}(x)$. This will be applied in
a future publication to
the problem of laser-induced desorbtion, where the resulting temperature
dependent
friction coefficient is used in the Langevin equation.\cite{laserteori}

In conclusion we have outlined a formalism that provide a theoretical
foundation of the DIMET point of view of desorbtion and applied this to
the non-equilibrium Eigler experiment.

\begin{figure}
\caption{The effective temperature as a function of the applied voltage
$eV$. The absolute ratio $\alpha / \eta$ is
estimated roughly using the ratio, Xe vibrational- to
electronic life-time, $\tau_{el}/\tau_{ph}\approx 100$, and
$\omega_0\approx 3 meV$.    }
\label{fig1:teff}
\end{figure}

\begin{figure}
\caption{The transfer rate, $\tau^{-1}$, as a function of junction voltage,
 for
$T=0K\;,4K\;,6K$ and $8K$.   }
\label{fig2:rate}
\end{figure}


\begin{references}
\bibitem{langevin}
Taking the classical limit, i.e. $kT$ much bigger than the frequencies of
the phonon-environment, $\eta$ will be the usual phenomenological
damping constant known from the classical Langevin equation,
$M\ddot x + \eta \dot x + V'(x)=F(t)$,
with $\langle F(t)F(t')\rangle=2\eta kT\delta(t-t')$
and $\langle F(t)\rangle = 0$.
\end{references}
\end{document}